%% file: main.tex
\documentclass[10pt,sigconf,letterpaper, nonacm]{acmart}

\AtBeginDocument{%
  }

\setcopyright{acmcopyright}
\copyrightyear{2022}
\acmYear{2022}
\acmDOI{123X567.123X567}

\acmConference[IMC '99]{ACM Internet Measurement Conference}{February 30, 1999}{World}
\acmPrice{15.00}
\acmISBN{978-1-4503-XXXX-X/22/10}

\usepackage{multirow}
\usepackage{booktabs}
\usepackage{pifont}
\usepackage{csquotes}
\usepackage{longtable}
\usepackage{supertabular}

\usepackage[breakable, skins]{tcolorbox}
\usepackage{adjustbox}

\tcbset{arc=0mm,size=fbox,attach title to upper={\ ---\ },coltitle=black}

\input{Defs}

\begin{document}

\title{Ethical and Privacy Considerations with Location Based Data Research}

\author{Leonardo Tonetto}
\affiliation{%
  \institution{Technical University of Munich}
  \streetaddress{Boltzmannstrasse 3}
  \city{Garching bei M\"{u}nchen}
  \country{Germany}}

\author{Pauline Kister}
\affiliation{%
  \institution{Technical University of Munich}
  \streetaddress{Boltzmannstrasse 3}
  \city{Garching bei M\"{u}nchen}
  \country{Germany}}

\author{Nitinder Mohan}
\affiliation{%
  \institution{Technical University of Munich}
  \streetaddress{Boltzmannstrasse 3}
  \city{Garching bei M\"{u}nchen}
  \country{Germany}}
\email{mohan@in.tum.de}

\author{J\"{o}rg Ott}
\affiliation{%
  \institution{Technical University of Munich}
  \city{Garching bei M\"{u}nchen}
  \country{Germany}
}
\email{ott@in.tum.de}


\renewcommand{\shortauthors}{Tonetto et al.}

\begin{abstract}
    \input{Abstract}
\end{abstract}


\maketitle

\input{article}


\balance{}%
\bibliographystyle{ACM-Reference-Format}
\bibliography{bibliography}




\end{document}

%% file: Defs.tex
\newcommand\Npapers{118}
\newcommand\Ndatasets{149}

\newcommand\NIRB{24}
\newcommand\PIRB{16.1}

\newcommand\NreportRows{81}
\newcommand\PreportRows{54.4}
\newcommand\Nstudydata{37}
\newcommand\Pstudydata{24.8}
\newcommand\Nshared{16}
\newcommand\Pshared{10.7}

\newcommand\NuidChanged{9}
\newcommand\PuidChanged{6.0}
\newcommand\Nhomeused{47}
\newcommand\Phomeused{37.3}
\newcommand\Nhome{126}
\newcommand\Phome{84.6}
\newcommand\Nplacesused{61}
\newcommand\Pplacesused{44.5}
\newcommand\Nplaces{137}
\newcommand\Pplaces{91.9}
\newcommand\Nscheduleused{61}
\newcommand\Pscheduleused{41.5}
\newcommand\Nschedule{147}
\newcommand\Pschedule{98.7}
\newcommand\Nsocialused{37}
\newcommand\Psocialused{25.5}
\newcommand\Nsocial{145}
\newcommand\Psocial{97.3}

\newcommand\Npreproc{10}
\newcommand\Ppreproc{6.7}
\newcommand\Npreprocmeth{6}
\newcommand\Ppreprocmeth{60.0}
\newcommand\Npreprocwho{9}
\newcommand\Ppreprocwho{90.0}
\newcommand\NreportFilteredRows{19}
\newcommand\PreportFilteredRows{12.8}
\newcommand\ProwsDiscarded{50.4}
\newcommand\PsubjectsDiscarded{62.1}
\newcommand\NreportFilteredSubjects{43}
\newcommand\PreportFilteredSubjects{28.9}
\newcommand\NethConsent{19}
\newcommand\PethConsent{12.8}

%% file: Abstract.tex
Networking research, especially focusing on human mobility, has evolved significantly in the last two decades and now relies on collection and analyzing larger datasets. The increasing sizes of datasets are enabled by larger automated efforts to collect data as well as by scalable methods to analyze and unveil insights, which was not possible many years ago. However, this fast expansion and innovation in human-centric research often comes at a cost of privacy or ethics.
In this work, we review a vast corpus of scientific work on human mobility and how ethics and privacy were considered. We reviewed a total of \Npapers{} papers, including \Ndatasets{} datasets on individual mobility.
We demonstrate that these ever growing collections, while enabling new and insightful studies, have not all consistently followed a pre-defined set of guidelines regarding acceptable practices in data governance as well as how their research was communicated.
We conclude with a series of discussions on how data, privacy and ethics could be dealt within our community.

%% file: article.tex


\section{Introduction}\label{sec:intro}

Understanding human mobility has been a quest of scientific research for many years, but it was not until the popularization of smart phones and its embedded sensors that we could deeply study it. The field itself has grown tremendously in the past two decades. From survey-based studies with limited number of subjects, sensing human movements through mobile systems enabled large scale studies benefiting the design of better communication protocols, urban policies and containment of infectious diseases. As the research problems and scope in the field evolved and grew, so did the data collection methods utilizing new sensing technologies, multiple experiment sources, number of involved subjects, etc. As such, the human mobility datasets have become larger and more diverse, more robust and complex methods and models were devised to study these sets, and a vast amount of work has revealed potential privacy implications with such data (\textit{e.g.}, de-anonymization methods~\cite{DeMontjoye2013,Wang2018,Freudiger2012}). 

To handle the growing ethical concerns from research studies involving human subjects, outlets for scientific articles are increasingly requiring ethical statements from authors at the time of submission itself. For example, venues associated with ACM, such as SIGCOMM (\textit{e.g.}, IMC, HotNets, CoNEXT) and SIGMOBILE (\textit{e.g.} MobiCom, MobiSys, MobiHoc) as well as workshops on the topic require authors to refer to their principles and code of ethics~\cite{belmont, acm-guideline} and submit a statement accompanying the article explicitly mentioning any ethical concerns raised by the study and steps undertaken by authors to mitigate them~\cite{10.1145/2793013}.
However, most of these policies are still evolving and ethical issues from published scientific studies and datasets have still not been eradicated~\cite{Partridge2016,Thomas2017}.
We believe that the primary contributor to this is the metamorphosis in the definition of \enquote{ethics} in research as the views around data ownership in collection evolve with technological advancements.
However, untimely discovery of ethical issues within research not only jeopardizes the privacy of involved subjects but can cause embarrassment to involved researchers through dataset redactions, erratas (sometimes retractions~\cite{minessota}) along with widespread clarifications.  

Despite serious implications, no recent study has reviewed and discussed how research articles, especially focusing on network research, used personal data of its subjects whilst dealing with privacy and ethics. 
In this paper, we plug this gap in research and focus on research on individual mobility data -- as those have been extensively used in recent networking research (\textit{e.g.},~\cite{Lutu2020,Wang2019,Heiler2020}) and arguably are a source of data the general public (outside of the research community) relates to most (\textit{e.g.},~\cite{NYTCovid2020}). 
In this article, we do not aim at reviewing the latest methods for studying human mobility, but rather we reexamine how individual location data was used in those studies, in light of recent developments in data regulations and ethics, along with the different properties of the data used. 
To this purpose, we discuss \textbf{(1)} the various types of datasets used for mobility research, \textbf{(2)} recent developments in data de-/anonymization that may shape public opinion on ethical concerns involving different data types, \textbf{(3)} the current state for regulations in certain countries for the use of individual location data, \textbf{(4)} how human mobility papers have treated this matter over the years, and finally \textbf{(5)} we discuss approaches future human mobility research could follow to establish a sustainable environment for both researchers and subjects.

To the best of our knowledge, ours is the first work to take a broader, multi-faceted look at the evolving state of human mobility research through ethical lenses.
We believe that the observations and discussions raised in this work can extend to other fields of research, wherever personal information of human subjects is involved.
As we point out later in this paper, we believe that more could be done to unify and standardize best practices on data governance as well as how research results should be communicated. 
The relevance of this topic lies in (1) protecting the privacy of those subjects being studied, and (2) preserving the trust of the general public with such research in a preventive manner, allowing for even more studies to be published.


\section{Related Work}\label{sec:relatedwork}

In recent years, a series of scientific papers have discussed different aspects of ethics in networking research. While Partridge \textit{et al.}~\cite{Partridge2016} advocated for an ``\emph{ethical considerations}'' section in all network measurements papers, the work by van der Ham \textit{et al.}~\cite{Ham2017Ethics} and Thomas \textit{et al.}~\cite{Thomas2017} reviewed the use of datasets in network measurements research. The work by van der Ham \textit{et al.}~\cite{Ham2017Ethics} discussed the ethical implications of four case studies where publicly available data contained more information than expected, compromising the privacy of their subjects. The work by Thomas \textit{et al.}~\cite{Thomas2017} reviewed the ethics of using datasets of illicit origins for research. Their work discussed the limitations regarding the lack of informed consent in such studies whenever humans are part of the leaked data. Furthermore, they present a series of case studies where leaked data were used in security related scientific papers. The authors present and rebut some commonly used arguments to justify the use of such data, and conclude that both researchers and scientific outlets should take a more responsible and ethical approach towards researching on stolen data. On the same topic, Ienca \textit{et al.}~\cite{ienca2021ethical} reviewed the use of hacked data in machine learning papers, suggesting pondering benefits and risks of each dataset while being clear about the means through which data were obtained and how researchers dealt with personal information, if present. From a different angle, we discuss how datasets legally obtained were used, while containing personal data and how these were communicated in their published manuscript.

The ethical implications of re-using data as well as reproducibility of results have also been addressed. Bot{\'{e}} \textit{et al.}~\cite{Bote2019} discussed the ethical and privacy problems associated with the re-usability of research data, arguing for its benefits regarding reproducibility and further research results from an already existing set. Furthermore, reviewing a series of Internet measurement papers, a recent study by Demir \textit{et al.}~\cite{Demir2022} reviewed the reproducibility of methods used in those papers, along with observations about the content of these papers. The authors observe that almost two-thirds of the analyzed papers (N=117) do not provide an ethics section, in spite of mostly focusing on security and privacy work. We make similar observations for papers using human mobility data, which is arguably a topic likely to reach non-scientific audiences. Additionally, we discuss other aspects related to privacy, such as IRB checks, consent

Furthermore, while Iqbal \textit{et al.}~\cite{Iqbal2019} reviewed the progress and citations of ACM SIGCOMM papers over recent decades, an early study by Kurkowski \textit{et al.}~\cite{Kurkowski2005} discussed how various methods for simulating mobility were used in the early 2000s. In this work, we include a survey covering over 20 years of research and how mobility datasets have evolved.

\section{Human Mobility Datasets}\label{sec:datasets}

Individual human mobility datasets may be collected from multiple sources, but are fundamentally described in the same way. They capture the historical presence of a person (or a handheld smart device) at a given point in space and time. Next we present a logical separation of the main classes of mobility data, with its main characteristics summarized on Table~\ref{tab:datasets}.

\begin{table*}[]
\caption{Summary of data types used for human mobility studies.}
\label{tab:datasets}
\begin{tabular}{@{}ccccc@{}}
\toprule
\textbf{Class} &
  \textbf{Source} &
  \multicolumn{1}{c}{\textbf{Location Accuracy}} &
  \multicolumn{1}{c}{\textbf{Sample Trigger}} &
  \multicolumn{1}{c}{\textbf{Area Covered}} \\ \midrule
\multirow{3}{*}{\textbf{\begin{tabular}[c]{@{}c@{}}Communication\\ Infrastructure\end{tabular}}} &
  Telco &
  100 m - 10 km~\cite{Pappalardo2021}&
  \begin{tabular}[c]{@{}l@{}}CDR: calls\\ XDR: calls+data\\ CPR: cell signalling\end{tabular} &
  city, country \\ \cmidrule(l){2-5} 
 &
  CCR &
  room, building~\cite{di2018sequences} &
  user purchase &
  city, country \\ \cmidrule(l){2-5} 
 &
  \multirow{2}{*}{Wi-Fi} &
  \multirow{2}{*}{room, building~\cite{Chaintreau2007}} &
  de-/association &
  campus, city \\ \cmidrule(r){1-1} \cmidrule(l){4-5} 
\multirow{3}{*}{\textbf{Experiments}} &
   &
   &
  + active scanning &
  any \\ \cmidrule(l){2-5} 
 &
  GPS sensor &
  10 m - 100 m~\cite{Stopczynski2014} &
  phone setting &
  any \\ \cmidrule(l){2-5} 
 &
  Bluetooth &
  10 m - room~\cite{Stopczynski2014} &
  phone setting &
  any \\ \midrule
\multirow{2}{*}{\textbf{\begin{tabular}[c]{@{}c@{}}Internet Based\\ Services\end{tabular}}} &
  OSN &
  10 m - 100 m~\cite{Jurdak2015} &
  content post &
  any \\ \cmidrule(l){2-5} 
 &
  Web &
  100 m - 10,000 km~\cite{Gouel2021} &
  page/app interaction &
  any \\ \midrule
\textbf{\begin{tabular}[c]{@{}c@{}}Public Transport\\ Infrastructure\end{tabular}} &
  SmartCard &
  0 m - 10 m~\cite{Sun2013Understanding} &
  get on/off transport &
  city \\ \bottomrule
\end{tabular}
\end{table*}

\subsection{Communication Infrastructure based}

This source of data relies on existing communication infrastructure to sense the presence of a subject. This presence is inferred by records (or logs) created when a subject's mobile device communicates with a point of access, which could be of a Wi-Fi network or of a mobile cellular network~\cite{Blondel2015}. While Wi-Fi setups are often limited to confined areas, such as companies or universities~\cite{Chaintreau2007,Alipour2018}, they offer room-level accuracy for their locations. Mobile cellular networks on the other hand may cover entire countries but have their accuracy between 100s meters to 10s kilometers~\cite{Gonzalez2008} with call detail records (CDR). In both cases, the availability of location records may be a function of the activity of the user, such as when an incoming/outgoing phone call happens or a device associates to an access point upon arriving to a new area. Additionally, certain network settings allow location records to be captured whenever \emph{any} data or even signaling events happen between the network and a subject's device (\textit{e.g.}, eXtended Detail Records (XDR) and control panel records (CPR)~\cite{Pappalardo2021}). For such, subjects often agree on \emph{terms and services} for the use of the offered infrastructure, and the collection of location data can be seen as a byproduct of this interaction. Alternatively, Credit Card Records (CCR) can also provide rich information about its user's whereabouts whenever a card is used for payment at a physical store~\cite{di2018sequences}.

\subsection{Designed Experiments}

Another class of human mobility data source originates from designed experiments, in which a phone app or pre-configured devices are distributed among subjects (\textit{e.g.}, \cite{Eagle2006,Stopczynski2014}). These types of effort often provide the highest level of flexibility and uniformity in how data are sampled, at times also providing data from other non-location sensors, such as accelerometers. These extra readings enable a higher accuracy in segmenting events, such as the duration of stops. However, as these studies are based on recruiting people, and given its costly setup and lack of secondary benefits for subjects (such as Internet access through a Wi-Fi access point), cohorts are limited in their size when compared to communication infrastructure based efforts~\cite{Stopczynski2014}. Location data collected through these studies often include continuous GPS coordinates, Bluetooth (BT) and Wi-Fi scans of nearby devices.

\subsection{Internet Based Services}

Another infrastructure provider that captures location data are those enabling the exchange of information on the Internet. Web services, such as online social networks (OSN)~\cite{Jurdak2015}, search engines~\cite{West2013} and map services~\cite{West2013} provide online users with services while also logging their physical location, for example by means of geotagged posts or geolocating devices based on their IP address~\cite{Callejo2022,Gouel2021}. Unlike the sources previously discussed, the sampling of location records by web services is highly dependent on how often a subject interacts with those services, leading to a skewed availability of data per device. Alternatively, location data may be captured in the background but only if allowed by the user (\textit{e.g.},~\cite{AppleLoc2022, AndroidLoc2022}).
Additionally, these web services can also capture extra features, such as the content of what is being searched or posted as well as the social graph of their users, which can be used to further enrich any analysis being made. Similar to the communication infrastructure, subjects agree to \emph{terms and conditions} for that service that states their location data will be logged and may (or may not) be used for further studies.

\subsection{Public Transport Infrastructure}

Smart Cards have replaced old paper-based ticketing system in most modern public transportation systems in large metropolitan areas~\cite{Sun2013Understanding}. They provide an integrated and automated way for passengers to pay for transport rides as well as manage different pricing schemes (\textit{e.g.}, senior citizens or students discounts). Users are required to present their smart cards before starting a ride, for example when entering a subway station or boarding a bus, and in some cases the same is expected when alighting. In this way, the system records the timestamps of discrete location points a subject has been. As human mobility tends to produce repeatable patterns~\cite{Song2010}, location data from public transport tends to be consistent and homogeneously through time, at least until a global pandemic changes how people move. Similar to all infrastructure-based sources, subjects agree to the \emph{terms and services} of using a smart card for their corresponding transport system.
Additionally, smart cards have also been used to trace the behavior of students in a university campus, logging various activities and services used by students~\cite{Lu2022}. Other examples of mobility data gathered from public transport are shared bikes~\cite{KALTENBRUNNER2010455}, taxis~\cite{TANG2015140}.

\vspace{0.15cm}

Given the various data sources discussed above, we now turn to a set of observations drawn on the scales of the papers surveyed and their respective datasets.

\section{Measurement Methodology}\label{sec:scales}

\begin{figure}[t]%
\centering
\includegraphics[width=0.99\columnwidth]{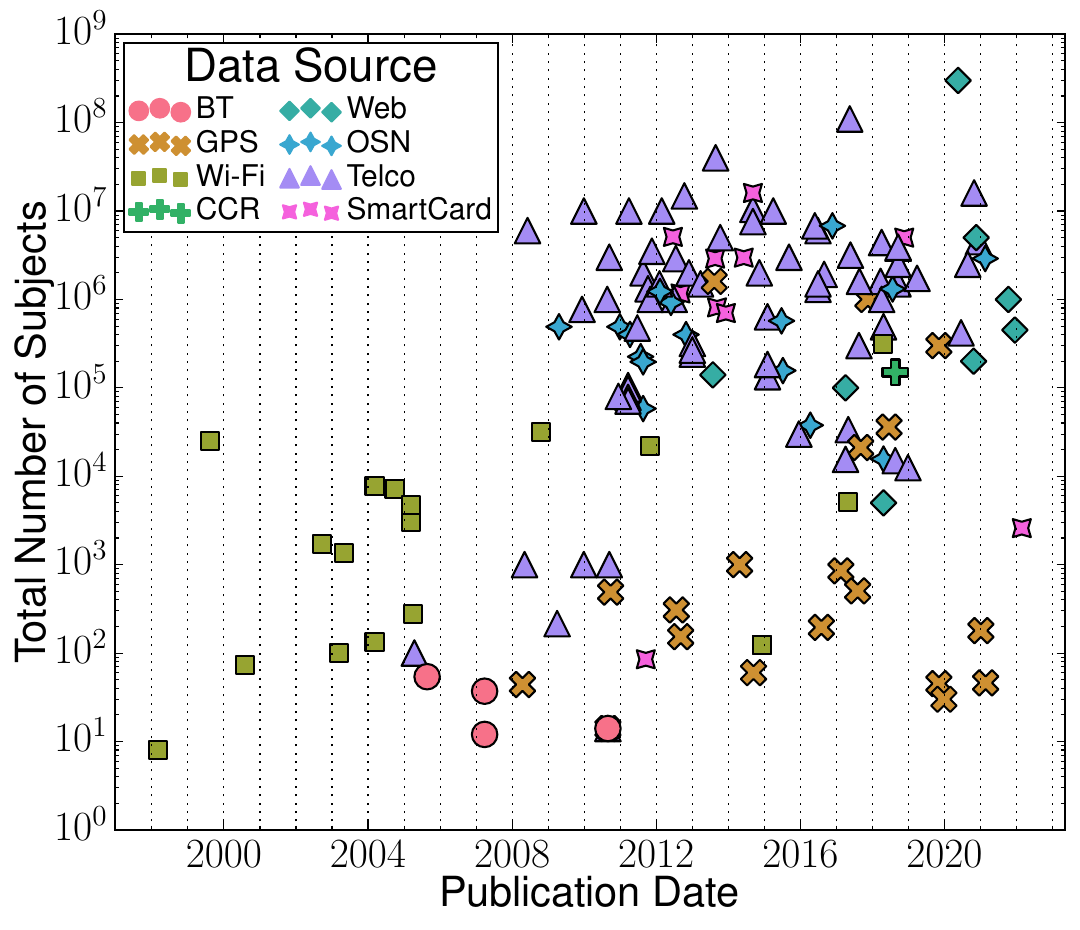}
\caption{Size of datasets per publication date.}\label{fig:scaletime}
\end{figure}

\subsection{Literature Selection}
\emph{Clarity} and \emph{justifiable choices} are of paramount importance when personal human data are used as an input to a research study. To understand how individual mobility data has been utilized in research till date, we conduct a survey on most relevant outlets for computer science with special focus on networking research venues (\textit{e.g.}, IMC, Mobicom, INFOCOM) and journals involving general sciences (\textit{e.g.}, PNAS, Nature, Science). Within this dataset, we filter for works that used human mobility dataset through keywords such as \emph{human mobility} and \emph{individual location data}, as well as including their respective relevant citations.
We further filtered studies where authors clearly stated that they had access to some form of user identifier (anonymized or not) and multiple spatial points per subject (\textit{e.g.}, geographical coordinates, points of interest) in the description. 
To limit the scope of this review, we explicitly excluded research from \textit{Transportation} and \textit{Geographic Information Systems} fields. 
We also did not consider any \textit{car/vehicle} dataset, as those often involved taxi drivers or cars with multiple owners (\textit{i.e.}, encompassing more than one subject). 

We performed this survey across the past 24 years (1998-2022) resulting in \textbf{\Npapers{}} research articles.
%
Of these \Npapers{} papers we surveyed, we identified a total of \textbf{\Ndatasets{}} datasets with varying properties based on the description given in their respective papers.
We categorized the papers in \emph{six} different buckets based on their method for estimating locations of involved subjects, specifically bluetooth (\emph{BT}), \emph{GPS}, \emph{WiFi}, credit card records (\emph{CCR}), web browser (\emph{web}), online social networks (\emph{OSN}), telecommunication networks (\emph{telco}) and public transport smartcards (\emph{smartcard}).
We observed that, thanks to the advancements in data collection methods, the collected sizes of mobility data has grown exponentially over the years (see figure~\ref{fig:scaletime}). Furthermore, we also observed that recent studies on the subject tend to incorporate more than one data sources to collect information of its subjects, as discussed in the previous section -- further expanding the involved dataset sizes, both in number of participants and the amount of collected data.

\subsection{Scale of Selected Studies}
\begin{figure}[t]%
\centering
\includegraphics[width=\columnwidth]{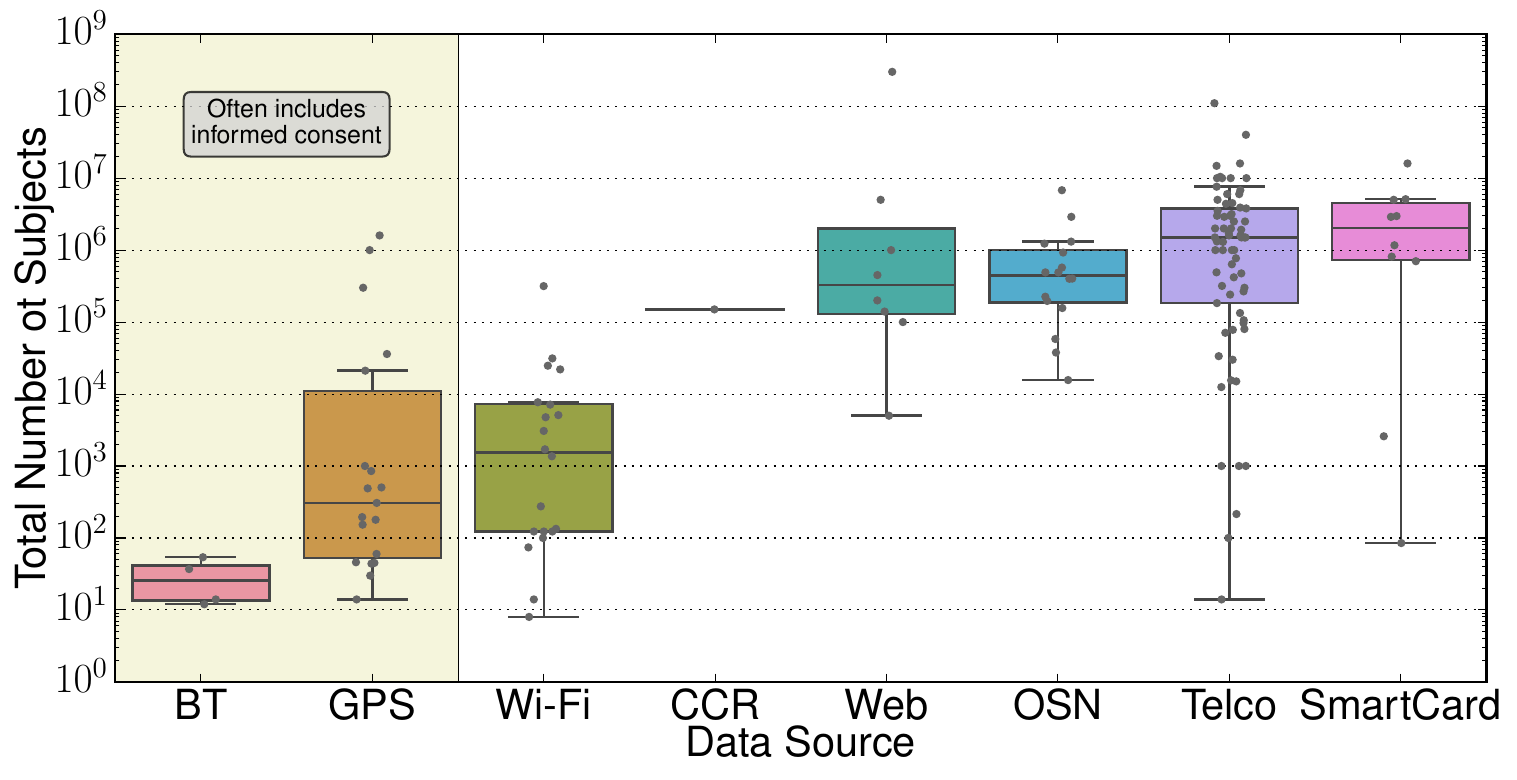}
\caption{Size of datasets per source.}\label{fig:sourcesize}
\end{figure}

\begin{figure}[t]%
\centering
\includegraphics[width=\columnwidth]{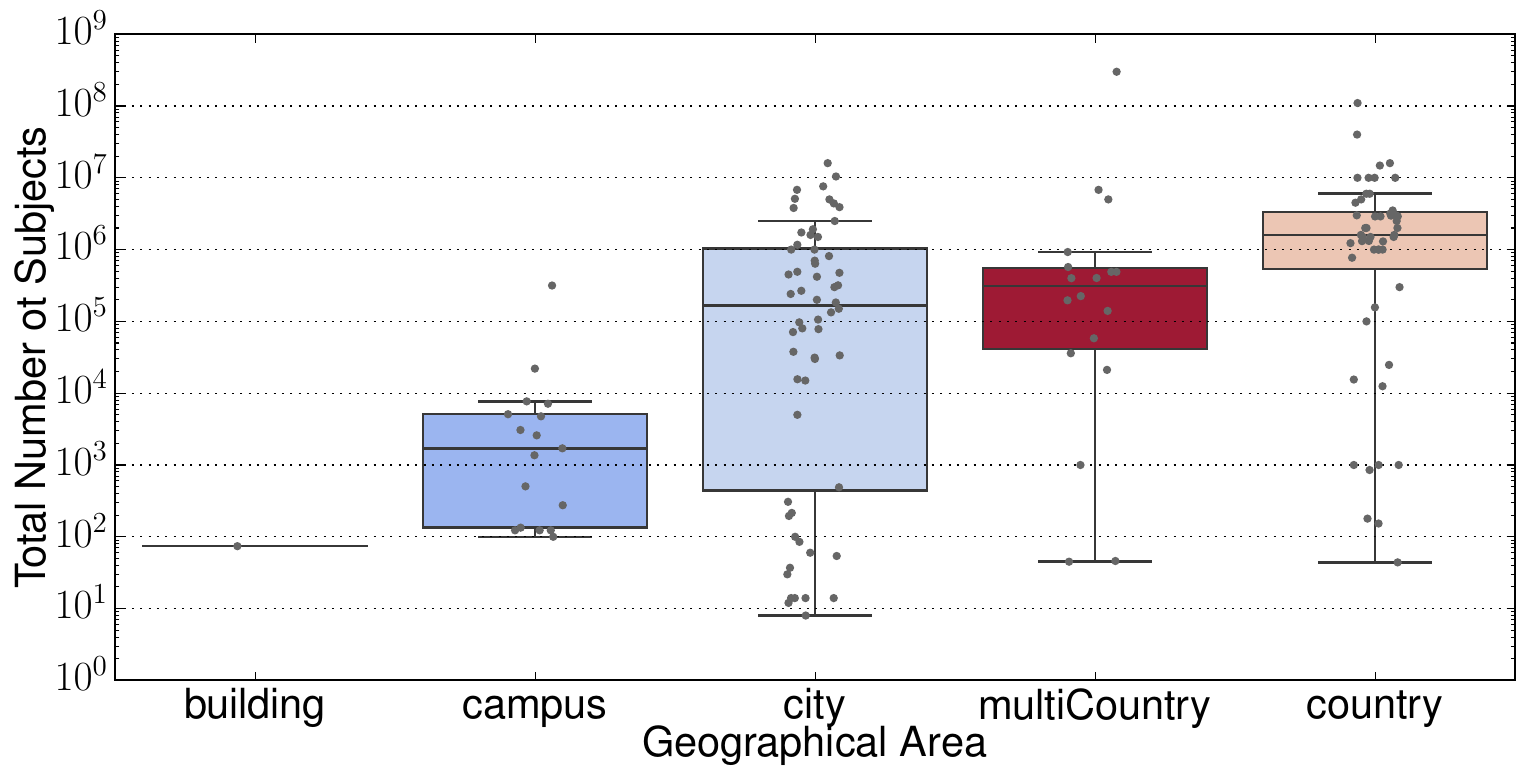}
\caption{Size of datasets per geographical area/type coverage.}\label{fig:geosize}
\end{figure}

With a growing deployment of Wi-Fi networks at universities and company facilities, logs from these setups have allowed researchers to obtain reliable mobility data from hundreds and even thousands of residents, including from their dormitories. While several mobility models were built with such data (\textit{e.g.},~\cite{BARBOSA20181}), these were mostly limited to workdays and covered only a small geographical area. To overcome that limitation, the late 2000s and early 2010s saw the emergence of a wide variety of mobility data sources, with even larger pool of subjects and, in a few cases, even covering multiple countries. From our analysis we also observed that while the number of subjects increased over time and larger data collections periods were used, although those did not correlate with the cohort sizes. Additionally, Figure~\ref{fig:sourcesize} depicts the distribution of dataset sizes per data source, further highlighting the disparities in cohort sizes between studies that often provide informed consent or not, as will be discussed later in this article. Unsurprisingly, sets covering larger areas also tend to include more users, as shown in Figure~\ref{fig:geosize}. However as multiple countries are only included in OSN and Web collections (see their limitations above) those do not necessarily have the largest groups.

We also find that a majority of articles tend to not properly describe the properties of their dataset well within their text. Surprisingly, almost half of the datasets (\NreportRows{}, or \PreportRows{}\%) within the selected articles did not report the number of records (or table rows) used in the study. For those which reported those numbers, we also observed a similar growth trend for overall number of rows in datasets over the years. Only one in eight of the total datasets (\NreportFilteredRows{}, or \PreportFilteredRows{}\%) reported the number of rows used (\textit{i.e.}, after initial filtering), and for those which reported values, on average half (\ProwsDiscarded{}\%) of the total rows were discarded. This filtering is typically done to reduce temporal sparsity without significantly reducing temporal resolution. Similarly for the number of subjects available for each study after filtering, one in four (\NreportFilteredSubjects{}, or \PreportFilteredSubjects{}\%) report total and filtered values where on average nearly two thirds (\PsubjectsDiscarded{}\%) of subjects are discarded. This filtering, while necessary for a homogenized set of subjects over time, may also introduce biases to the results, which should also be discussed. In Section~\ref{sec:ethicalconsiderations} we discuss how the papers we surveyed often had access to more information about their subjects than it was actually needed for their studies, such as \textit{home} location and their social graph.

\begin{figure}[t]%
\centering
\includegraphics[width=\columnwidth]{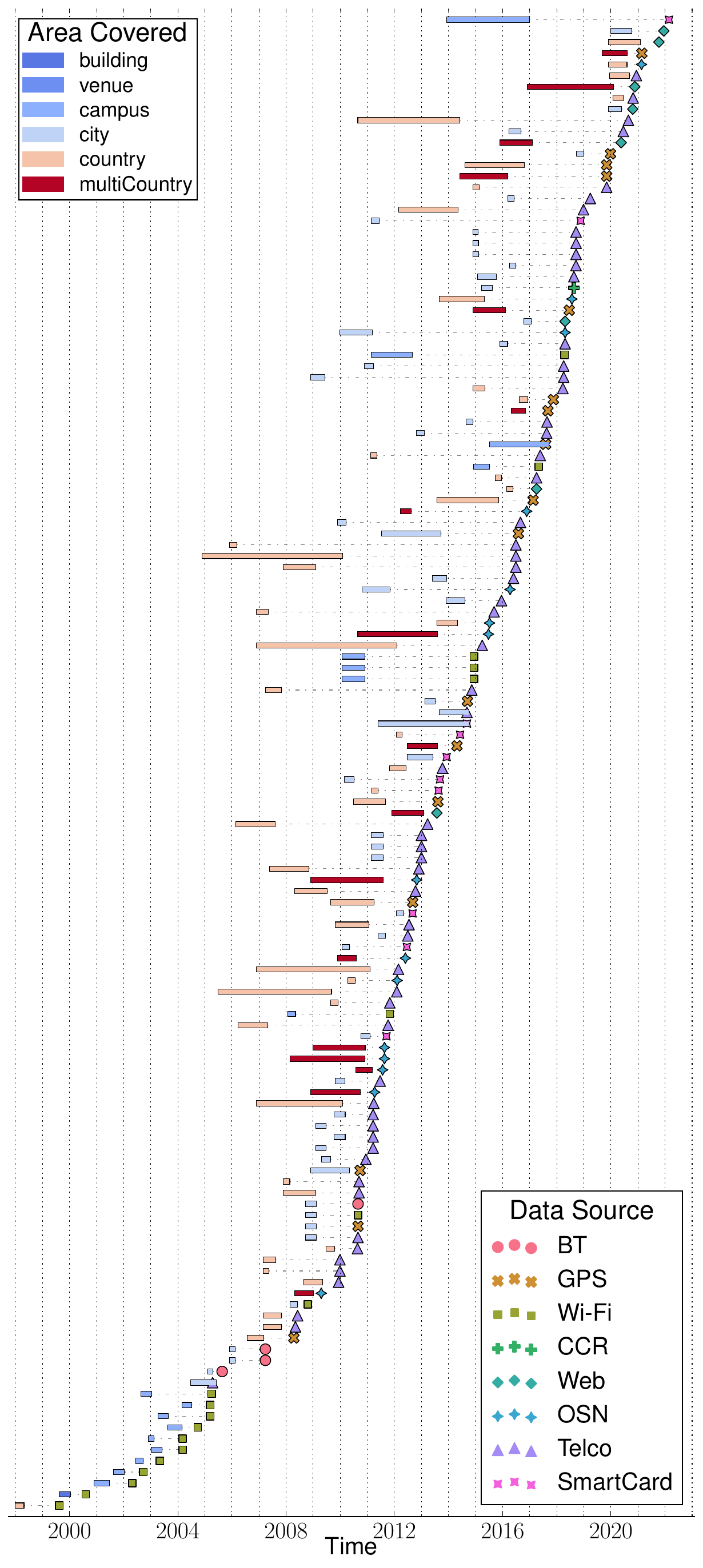}
\caption{Data source, sampling duration and area type per dataset. Entries sorted by date their papers were published.}\label{fig:sampleduration}
\end{figure}

Our analysis on the scale of mobility datasets provides context to the increasing magnitude in the number of subjects and rows as well as in geographical extent of measurement studies in the field over the years.
While the scales of datasets used for such studies has grown significantly, so has the general public's awareness of possible privacy issues related to those data~\cite{Alrayes2016}. Complementary to this awareness, as new research on anonymization methods became popular, new de-anonymization approaches have followed suit. Next, we review the literature on such methods as they support our understanding of how future research on human mobility ought to follow.

\section{The Private Data Conundrum}\label{sec:privatedata}


In 2006, AOL shared a sample of historical queries from over 650,000 users for research purposes. The identities of these users were altered in order to protect their privacy, but shortly after its release a series of users were \emph{de-anonymized} making the headlines of newpapers (\textit{e.g.}, \cite{NYT2006,searchids2006}). Furthermore, even data not released on purpose may be leaked through security breaches (\textit{e.g.}, \cite{BBC2017}). Consequently, these events populate public opinion about data security and privacy, leading to new or modified legislation as well as ethical guidelines. These new directions finally shape recent network research by limiting what can be done when personal data are used, or at least it should.

To tackle these constraints when dealing with personal data, researchers have paid significant attention to methods for both anonymizing private personal records, simultaneously to de-anonymize it. These works often aim at raising awareness of possible weaknesses in identity protection, and how to provide minimal guarantees to one's privacy, respectively.

Even though we do not analyze the full life-cycle of datasets, in several occasions sets were either made public or at least used in further studies carried by the original research group. On that topic, Allman and Paxson addressed the ethical concerns of releasing and using data, including privacy protection~\cite{AllmanP07}. They emphasize that raw datasets containing identifiable information should not be shared publicly, and research should prioritize preserving the privacy of individuals. Additionally, the authors advise against trusting anonymization approaches as deanonymization counter-attacks quickly become available.

We now provide a brief review of some of such methods as those are often cited, as we observe, as reasons for an unconditional care whenever individual data are used.

\subsection{Location-based data de-anonymization techniques}

When sharing individual data, personal attributes which can \emph{directly} identify any subject (\textit{e.g.}, name or phone number) may be changed to \emph{pseudonyms} in order to protect their privacy. These steps, however, yield only partial privacy protection as highlighted by a plethora of studies on de-anonymization attacks~\cite{fung2010privacy,fiore2020privacy}. Following the taxonomy of Fung \textit{et al.}~\cite{fung2010privacy} and Fiore \textit{et al.}~\cite{fiore2020privacy}, we review some of these attacks which we later refer to when discussing individual location data in mobility research. The relevance of these studies to our current work lies in the observation that all papers we surveyed state that user identifiers were \emph{anonymized}, and in some cases that being the only privacy related content in the manuscript.

On a study on attacks through \emph{record linkage}, in which an attacker has a database containing (quasi-)similar location records from their victims, De Montjoye \textit{et al.}~\cite{DeMontjoye2013} showed how any two location records could de-anonymize 50\% of subjects and how four locations could increase this trajectory uniqueness up to 95\%, from a set of 1.5 million persons. However, a comprehensive study by Wang \textit{et al.}~\cite{Wang2018} showed that former methods actually underperformed when applied to real-world large scale data, reporting a de-anonymization hit rate lower than 20\% in all cases. Wang \textit{et al.} then finally proposed a Markov-based model that achieves a 40\% success rate in linking records from heterogeneous databases.

Another relevant surface of attack is through a side-channel, such as a victim's profile. By using both \emph{home} and \emph{work} information of victims, Freudiger \textit{et al.}~\cite{Freudiger2012} show how an adversary can uniquely identify individuals in large datasets with up to 90\% accuracy. Furthermore, knowledge on the whereabouts of one's friends, or her \emph{social graph}, may reveal relevant information about her current location. Using data from online social networks, Sadilek \textit{et al.}~\cite{Sadilek2012} proposed an unsupervised method, achieving 57\% and 77\% accuracy in predicting a victim's location given information from two and nine of her friends, respectively. Similarly with a supervised learning approach, the authors obtained accuracies of 77\% and 84.3\%, respectively. Using a different approach and a set of smaller but denser networks, Srivatsa \textit{et al.}~\cite{Srivatsa2012} reported that up to 80\% of subjects' identity could be recovered by matching social graphs links with increased physical proximity.

Taken together, these attacks are vivid examples of how individual location data can uniquely identify a person and what some vectors for de-anonymization are. These threats may shape not only the contents of academic research but also the general public view on the harm such data impose to their individual privacy. To counteract these perils and uncertainties two alternatives are often presented: (1) stronger anonymization methods, and (2) ethical guidelines and regulations aimed at protecting a person's right for anonymity. We further discuss both of these point, next.


\subsection{Stronger anonymization for location-based data} \label{subsec:anon}

Given the threats to an individual's privacy brought by the de-anonymization attacks discussed above, a growing literature on anonymization methods has produced relevant results. Following the taxonomy of Fiore \textit{et al.}~\cite{fiore2020privacy}, anonymization techniques follow three main principles. (1) \emph{Indistinguishability}, in which the records of a subject must be indistinguishable from those of the same size-limited anonymity set, essentially removing the unicity from one's trajectory. A commonly used formal approach for indistinguishability is $k$\emph{-anonymity}~\cite{SWEENEY2002}. (2) \emph{Uninformativeness}, in which an adversary's knowledge about a victim must remain unchanged after accessing a dataset the victim is part of. This principle is generally achieved through differential privacy~\cite{Dwork2008}. (3) \emph{Mitigation}, in which preventive measures are taken to assure one's data privacy, without a well-define principle. Such principle is addressed with alterations in the original location data, aimed at discarding certain unique aspects of one's trajectories, however without providing formal guarantees.

On \emph{indistinguishability}, ensuring $k$\emph{-anonymity} has taken a multitude of solutions, such as: (a) \emph{spatiotemporal generalization}, in which spatial and temporal resolutions are reduced until traces from members of an anonymity set are indistinguishable. Achieving this goal, however, quickly reduces the quality of an anonymized location dataset as more trajectories are aggregated, as revealed by De Montjoye \textit{et al.}~\cite{DeMontjoye2013}. (b) \emph{Suppression}, in which points are removed from trajectories until their unicity is lost~\cite{terrovitis2008privacy}. (c) \emph{Spatial uncertainty}, in which $k$\emph{-anonymity} is relaxed to include entire trajectories that fall within a defined threshold, as demonstrated by Abul \textit{et al.}~\cite{Abul2008}. It is important to note that $k$\emph{-anonymity} has its own shortcomings, addressed by extensions such as $l$\emph{diversity} that enforces an extra diversity criteria per group~\cite{machanavajjhala2007diversity}.

On \emph{uninformativeness}, proposed solutions formally follow differential privacy in which consecutive similar queries to a dataset should return similar results, even if the trajectories of a subject have been added or removed from the set. This has been shown, for example, by Shokri \textit{et al.}~\cite{Shokri2012} who adapted differential privacy for location data. Other examples of differential privacy include altering the representation of the trajectory data using different data structures~\cite{chen2012differentially} as well as aggregating observations as probability distributions~\cite{gursoy2018differentially}.

On \emph{mitigation}, proposed solutions do not provide formal privacy guarantees and also often reduce the utility of the original location data. This is achieved through: (a) \emph{obfuscation}, in which noise is added to location data (\textit{e.g.}, \cite{duckham2005formal}). (b) \emph{Cloaking}, in which spatial or temporal resolutions are reduced (\textit{e.g.}, \cite{rossi2015spatio}). (c) \emph{Segmentation}, in which new pseudonyms are used for multiple segments that are artificially added to the data (\textit{e.g.}, Song \textit{et al.}~\cite{DBLP:conf/sigir/SongDB14}). (d) \emph{Swapping}, in which pseudonyms are randomly swapped between different trajectories (\textit{e.g.}, \cite{salas2020swapping}).


\smallskip
The observations in these studies remind us that replacing real identifiers by pseudonyms is not enough to protect one's privacy. A myriad and ever growing literature on de-anonymization approaches highlights not only the risks of dealing with personal data, but also the importance of stronger anonymization techniques. These protective methods, however, often degrade information present only in the original datasets. This inevitable drawback affects what studies are capable of doing, by eliminating important nuances, as well as whether or not data can be shared, and in which format.

\section{Ethical guidelines and regulations outlook} \label{sec:ethics} 

Given all the risks to one's privacy regarding location data, ethical guidelines as well as regulatory bodies have adapted their take on these topics. These efforts are of paramount importance as they aim at protecting subjects in circumstances where only hindsight can assure privacy, even if only partially. Ethical considerations are relevant in all stages of human mobility research, from the design of research questions, the analysis of data, the articles being reviewed and published as well as to the dissemination of the work that is either being done or that is already completed~\cite{franzke2020ethics}.

To safeguard individual's privacy, the European Union General Data Protection Regulation (GDPR) noticeably covers exclusively personal data. In the GDPR, personal data are defined with respect to the identifiability of an individual based on the data being regarded. However, if data are not personal or have been sufficiently anonymized they are no longer protected by the GDPR laws~\cite{british2017data}. As we previously discussed, there is a large body of research on how most methods of anonymization are flawed and do not guarantee lack of unicity. Therefore, human mobility research for often dealing with personal location data has to ensure not only laws are met, but that research is conducted and presented following up-to-date ethical guidelines. This would ensure public \emph{trust} and the continuity of this research area.

With a similar purpose, the California Consumer Privacy Act (CCPA)~\cite{CCPA2018} establishes a set of rules to protect the privacy of online users, with special regards to the sale of personal data. A fundamental difference between the CCPA and the GDPR concerns the origin of the data covered by each law. While the former covers data provided by the consumer the latter makes no distinction, as long as it is identifiable.

An inexorable aspect of privacy, both ethically and in various law instances, is the need for consent. From a subject's point of view, individuals have reported being more willing to consent with the use of their data if the proposed research had clear benefits, either personally or for society~\cite{wellcome2013summary}. That is, if one assumes a utilitarian view, where benefits may justify potential risks, subjects are more likely to agree in taking part in a study if either them or society may benefit from the results of the study. One issue here, we argue, is that unlike medical research, human mobility studies from computer or social scientists are less tangible. Therefore, drawing a line to what is or is not ethical becomes harder. Nevertheless, mobility research can still assure subjects when data are being collected, what analyses are being done, and what the overall impact might be as those elements define public opinion on whether or not their data should be used~\cite{royal2017data,bakir2015public}. We note, however, that ensuring this transparency between researchers and subjects is often referred in the surveyed papers as one of the main reasons \emph{no data} can be shared. As a result, the \emph{reproducibility} of results is impaired, which we further discuss in Section~\ref{sec:outlook}.

Another important facet of this complex problem is the ethical tradition of researchers and subjects' culture in different countries~\cite{franzke2020ethics}. Traditionally, US and UK scientists are more \emph{utilitarian} in which the value of the research being done justifies the potential ethical risks associated, whereas their European counterparts favor a more \emph{deontological} view, \textit{i.e.}, it favors moral values as well as rules before weighing and considering the outcome. Regarding subjects that in some studies may be from several countries, such as those using OSN data, their personal and cultural understanding and assumptions on privacy are \emph{ambiguous, contested and changing}~\cite{franzke2020ethics}. That is, given this vast combinatorial of cultural understandings and expectations regarding privacy and its associated risks, a cautious and preventive approach is undoubtedly beneficial in helping guide future scientific (and possibly commercial) efforts using personal location data.

There are, however, a series of steps human mobility research can follow in order to safeguard subjects' privacy as well as scientific and general public's trust on what is being studied.
Even though these actions often require additional intricate effort from researchers, we observe an increase in its adoption by publication outlets.
These include: (a) reviews from ethical review boards (or institutional review board, ERB/IRB), with these reviews/assessments being required to be independently verifiable (\textit{e.g.}, with a protocol number); (b) including comprehensive ethical statements, where the possible \emph{vulnerability} of and \emph{harm} to subjects is discussed, and what might have been the expectations of users of a studied service when interacting with it; (c) whether or not informed consent was given to the specific study being conducted, beyond the \emph{terms and services} of that platform providing the location data; and finally overall; and (d) what was done to minimize \emph{risk} for subjects, given all that has been discussed thus far. In the next section we review how, in the past, research papers have addressed some of the outstanding challenges human mobility faces regarding ethics, followed by a discussion on how we believe this should be done in the future.



\section{Ethical considerations in the literature} \label{sec:ethicalconsiderations}

\begin{table*}
\centering
\caption{Papers surveyed. \emph{Year} published, if \emph{Deliberate} privacy was considered, if an ethical \emph{Statement} was present, if an \emph{IRB} approval was present, if subjects gave \emph{Consent} to the exact terms of the study, and which data \emph{Sources} were used. Sources: W - Wi-Fi; T - Telco; B - Bluetooth; O - Online Social Networks; S - Smartcards; E - Web Services; C - Credit Card Records. (\ding{108} - Yes; \ding{109} - No)}
\label{tab:summary}
\fontsize{10}{0.5}\selectfont
\setlength{\tabcolsep}{2pt}
\begin{minipage}{0.475\textwidth}
\begin{tabular}{lcccccr}
\toprule
Paper & Year & Deliberate & Statement & IRB & Consent & Sources \\
\midrule
\cite{Lai1998} & 1998 & \ding{109} & \ding{109} & \ding{109} & \ding{109} & W \\
\cite{Tang1999} & 1999 & \ding{109} & \ding{109} & \ding{109} & \ding{109} & W \\
\cite{Tang2000} & 2000 & \ding{109} & \ding{109} & \ding{108} & \ding{109} & W \\
\cite{Hutchins2002} & 2002 & \ding{109} & \ding{109} & \ding{109} & \ding{109} & W \\
\cite{Kotz2005} & 2002 & \ding{109} & \ding{109} & \ding{109} & \ding{109} & W \\
\cite{Thajchayapong2003} & 2003 & \ding{109} & \ding{109} & \ding{109} & \ding{109} & W \\
\cite{Balazinska2003} & 2003 & \ding{109} & \ding{109} & \ding{109} & \ding{109} & W \\
\cite{Henderson2004} & 2004 & \ding{109} & \ding{109} & \ding{109} & \ding{109} & W \\
\cite{Chinchilla2004} & 2004 & \ding{108} & \ding{109} & \ding{109} & \ding{109} & W \\
\cite{Schwab2004} & 2004 & \ding{108} & \ding{109} & \ding{109} & \ding{109} & W \\
\cite{McNett2005} & 2005 & \ding{109} & \ding{109} & \ding{109} & \ding{109} & W \\
\cite{Eagle2006} & 2005 & \ding{109} & \ding{108} & \ding{109} & \ding{108} & T \\
\cite{Tuduce2005} & 2005 & \ding{109} & \ding{109} & \ding{109} & \ding{109} & W \\
\cite{Hui2005} & 2005 & \ding{108} & \ding{109} & \ding{109} & \ding{108} & B \\
\cite{Chaintreau2007} & 2007 & \ding{108} & \ding{109} & \ding{109} & \ding{108} & B \\
\cite{Rhee2008} & 2008 & \ding{109} & \ding{109} & \ding{109} & \ding{109} & G \\
\cite{Afanasyev2008} & 2008 & \ding{109} & \ding{109} & \ding{109} & \ding{109} & W \\
\cite{Gonzalez2008} & 2008 & \ding{109} & \ding{109} & \ding{108} & \ding{109} & T \\
\cite{Song2010} & 2009 & \ding{109} & \ding{109} & \ding{109} & \ding{109} & T \\
\cite{tatem2009use} & 2009 & \ding{109} & \ding{109} & \ding{109} & \ding{109} & T \\
\cite{Eagle2009} & 2009 & \ding{108} & \ding{109} & \ding{109} & \ding{108} & T \\
\cite{Crandall2009} & 2009 & \ding{109} & \ding{109} & \ding{109} & \ding{109} & O \\
\cite{Cranshaw2010} & 2010 & \ding{109} & \ding{109} & \ding{109} & \ding{108} & G \\
\cite{Crandall2010} & 2010 & \ding{109} & \ding{109} & \ding{109} & \ding{109} & O \\
\cite{Quercia2010} & 2010 & \ding{109} & \ding{109} & \ding{109} & \ding{109} & T \\
\cite{song2010modelling} & 2010 & \ding{109} & \ding{109} & \ding{109} & \ding{109} & T \\
\cite{Jensen2010} & 2010 & \ding{109} & \ding{109} & \ding{109} & \ding{109} & T, G, W, B \\
\cite{Phithakkitnukoon2010} & 2010 & \ding{109} & \ding{109} & \ding{109} & \ding{109} & T \\
\cite{Kang2012} & 2011 & \ding{109} & \ding{109} & \ding{109} & \ding{109} & T \\
\cite{Bagrow2011} & 2011 & \ding{109} & \ding{109} & \ding{109} & \ding{109} & T \\
\cite{Becker2011} & 2011 & \ding{108} & \ding{108} & \ding{109} & \ding{109} & T \\
\cite{Lathia2011} & 2011 & \ding{108} & \ding{108} & \ding{109} & \ding{108} & S \\
\cite{Scellato2011} & 2011 & \ding{109} & \ding{109} & \ding{109} & \ding{109} & O \\
\cite{Isaacman2011} & 2011 & \ding{108} & \ding{108} & \ding{109} & \ding{109} & T \\
\cite{phithakkitnukoon2011inferring} & 2011 & \ding{109} & \ding{109} & \ding{109} & \ding{109} & T \\
\cite{Phithakkitnukoon2011Out} & 2011 & \ding{109} & \ding{109} & \ding{109} & \ding{109} & T \\
\cite{Moghaddam2011} & 2011 & \ding{109} & \ding{109} & \ding{109} & \ding{109} & W \\
\cite{Cho2011} & 2011 & \ding{109} & \ding{109} & \ding{109} & \ding{109} & O, T \\
\cite{Cheng2011} & 2011 & \ding{109} & \ding{109} & \ding{109} & \ding{109} & O \\
\cite{Sadilek2012} & 2012 & \ding{109} & \ding{108} & \ding{109} & \ding{109} & O \\
\cite{Lathia2012} & 2012 & \ding{109} & \ding{109} & \ding{109} & \ding{109} & S \\
\cite{Csaji2012} & 2012 & \ding{109} & \ding{109} & \ding{109} & \ding{109} & T \\
\cite{blumenstock2012inferring} & 2012 & \ding{109} & \ding{108} & \ding{109} & \ding{109} & T \\
\cite{Noulas2011} & 2012 & \ding{109} & \ding{109} & \ding{109} & \ding{109} & O \\
\cite{Ranjan2012} & 2012 & \ding{109} & \ding{109} & \ding{109} & \ding{109} & T \\
\cite{Do2012} & 2012 & \ding{108} & \ding{109} & \ding{109} & \ding{108} & G \\
\cite{Allamanis2012} & 2012 & \ding{109} & \ding{109} & \ding{109} & \ding{109} & O \\
\cite{wesolowski2012quantifying} & 2012 & \ding{109} & \ding{109} & \ding{109} & \ding{109} & T \\
\cite{Sadilek2012Far} & 2012 & \ding{109} & \ding{109} & \ding{109} & \ding{108} & G \\
\cite{Williams2012} & 2012 & \ding{109} & \ding{109} & \ding{109} & \ding{109} & S \\
\cite{simini2012universal} & 2012 & \ding{109} & \ding{109} & \ding{109} & \ding{109} & T \\
\cite{lu2012predictability} & 2012 & \ding{109} & \ding{109} & \ding{109} & \ding{109} & T \\
\cite{Bhattacharya2013} & 2013 & \ding{109} & \ding{109} & \ding{109} & \ding{109} & S \\
\cite{Song2013} & 2013 & \ding{109} & \ding{109} & \ding{109} & \ding{109} & G \\
\cite{lu2013approaching} & 2013 & \ding{108} & \ding{109} & \ding{109} & \ding{109} & T \\
\cite{DeMontjoye2013} & 2013 & \ding{109} & \ding{109} & \ding{109} & \ding{109} & T \\
\cite{Sun2013Understanding} & 2013 & \ding{109} & \ding{109} & \ding{109} & \ding{109} & S \\
\cite{Becker2013} & 2013 & \ding{109} & \ding{108} & \ding{109} & \ding{109} & T \\
\cite{Ponieman2013} & 2013 & \ding{109} & \ding{109} & \ding{109} & \ding{109} & T \\
\bottomrule
\end{tabular}
\end{minipage}
\hspace{0.5cm}
\begin{minipage}{0.475\textwidth}
\begin{tabular}{lcccccr}
\toprule
Paper & Year & Deliberate & Statement & IRB & Consent & Sources \\
\midrule
\cite{West2013} & 2013 & \ding{109} & \ding{109} & \ding{109} & \ding{109} & E \\
\cite{Yuan2013} & 2013 & \ding{109} & \ding{109} & \ding{109} & \ding{109} & S \\
\cite{Vu2015} & 2014 & \ding{108} & \ding{108} & \ding{108} & \ding{109} & W \\
\cite{Wang2014Student} & 2014 & \ding{108} & \ding{108} & \ding{108} & \ding{108} & G \\
\cite{Stopczynski2014} & 2014 & \ding{108} & \ding{108} & \ding{108} & \ding{108} & G \\
\cite{deville2014dynamic} & 2014 & \ding{109} & \ding{109} & \ding{109} & \ding{109} & T \\
\cite{Zhang2014} & 2014 & \ding{109} & \ding{108} & \ding{109} & \ding{109} & T, S \\
\cite{sun2014efficient} & 2014 & \ding{109} & \ding{109} & \ding{109} & \ding{109} & S \\
\cite{Wang2015} & 2015 & \ding{109} & \ding{109} & \ding{109} & \ding{109} & T \\
\cite{pappalardo2015returners} & 2015 & \ding{109} & \ding{109} & \ding{109} & \ding{109} & T \\
\cite{lenormand2015human} & 2015 & \ding{109} & \ding{109} & \ding{109} & \ding{109} & O \\
\cite{Barbosa2015} & 2015 & \ding{109} & \ding{109} & \ding{109} & \ding{109} & T \\
\cite{toole2015coupling} & 2015 & \ding{109} & \ding{109} & \ding{109} & \ding{109} & T \\
\cite{Jurdak2015} & 2015 & \ding{109} & \ding{108} & \ding{108} & \ding{109} & O \\
\cite{Jiang2016} & 2016 & \ding{109} & \ding{109} & \ding{109} & \ding{109} & T \\
\cite{Deville2016} & 2016 & \ding{109} & \ding{109} & \ding{108} & \ding{109} & T \\
\cite{Vhaduri2016} & 2016 & \ding{108} & \ding{108} & \ding{108} & \ding{108} & G \\
\cite{Oliveira2016} & 2016 & \ding{109} & \ding{109} & \ding{109} & \ding{109} & T \\
\cite{Hristova2016} & 2016 & \ding{109} & \ding{109} & \ding{109} & \ding{109} & O \\
\cite{Thilakarathna2017} & 2016 & \ding{109} & \ding{109} & \ding{109} & \ding{109} & O \\
\cite{Thuillier2018} & 2017 & \ding{108} & \ding{109} & \ding{109} & \ding{109} & T \\
\cite{luo2017inferring} & 2017 & \ding{109} & \ding{108} & \ding{108} & \ding{109} & T \\
\cite{Vhaduri2017} & 2017 & \ding{108} & \ding{109} & \ding{109} & \ding{108} & G \\
\cite{Xu2017Trajectory} & 2017 & \ding{109} & \ding{109} & \ding{109} & \ding{109} & E, T \\
\cite{Zeng2017} & 2017 & \ding{109} & \ding{109} & \ding{109} & \ding{109} & G \\
\cite{Andone2017} & 2017 & \ding{108} & \ding{109} & \ding{108} & \ding{108} & G \\
\cite{alhasoun2017city} & 2017 & \ding{109} & \ding{109} & \ding{109} & \ding{109} & T \\
\cite{Jiang2017} & 2017 & \ding{109} & \ding{109} & \ding{109} & \ding{109} & T \\
\cite{Cao2017On} & 2017 & \ding{109} & \ding{109} & \ding{109} & \ding{109} & W \\
\cite{Cao2017} & 2017 & \ding{109} & \ding{109} & \ding{109} & \ding{109} & T \\
\cite{Alessandretti2017} & 2017 & \ding{108} & \ding{108} & \ding{108} & \ding{108} & G \\
\cite{alessandretti2018evidence} & 2018 & \ding{108} & \ding{108} & \ding{108} & \ding{108} & G \\
\cite{Wang2019} & 2018 & \ding{108} & \ding{109} & \ding{109} & \ding{109} & T \\
\cite{Xu2018} & 2018 & \ding{109} & \ding{109} & \ding{109} & \ding{109} & T \\
\cite{Wang2018Urban} & 2018 & \ding{109} & \ding{109} & \ding{109} & \ding{109} & O \\
\cite{Chen2018} & 2018 & \ding{109} & \ding{109} & \ding{109} & \ding{109} & T \\
\cite{Fang2018} & 2018 & \ding{109} & \ding{109} & \ding{109} & \ding{109} & T \\
\cite{Alipour2018} & 2018 & \ding{109} & \ding{109} & \ding{109} & \ding{109} & W \\
\cite{Sadri2018} & 2018 & \ding{109} & \ding{109} & \ding{109} & \ding{109} & T \\
\cite{Huang2018} & 2018 & \ding{109} & \ding{109} & \ding{109} & \ding{109} & S \\
\cite{Feng2018} & 2018 & \ding{109} & \ding{109} & \ding{109} & \ding{109} & O, E, T \\
\cite{di2018sequences} & 2018 & \ding{109} & \ding{109} & \ding{109} & \ding{109} & C, T \\
\cite{Teixeira2019} & 2019 & \ding{108} & \ding{109} & \ding{109} & \ding{109} & G, T \\
\cite{Chung2019} & 2019 & \ding{109} & \ding{109} & \ding{109} & \ding{109} & G \\
\cite{Tsubouchi2019} & 2019 & \ding{108} & \ding{109} & \ding{109} & \ding{109} & G \\
\cite{Wang2019Model} & 2019 & \ding{108} & \ding{108} & \ding{108} & \ding{109} & T \\
\cite{Heiler2020} & 2020 & \ding{108} & \ding{108} & \ding{109} & \ding{109} & T \\
\cite{Lutu2020} & 2020 & \ding{108} & \ding{108} & \ding{108} & \ding{109} & T \\
\cite{Giles2020} & 2020 & \ding{109} & \ding{109} & \ding{109} & \ding{109} & T \\
\cite{kraemer2020mapping} & 2020 & \ding{108} & \ding{108} & \ding{108} & \ding{109} & E \\
\cite{alessandretti2020scales} & 2020 & \ding{109} & \ding{108} & \ding{108} & \ding{109} & E \\
\cite{gauvin2020gender} & 2020 & \ding{109} & \ding{108} & \ding{108} & \ding{109} & T \\
\cite{yabe2020non} & 2020 & \ding{109} & \ding{109} & \ding{109} & \ding{109} & E \\
\cite{Yabe2022Early} & 2021 & \ding{108} & \ding{108} & \ding{108} & \ding{108} & E \\
\cite{Shida2021} & 2021 & \ding{109} & \ding{108} & \ding{109} & \ding{109} & E \\
\cite{Jarv2021} & 2021 & \ding{108} & \ding{109} & \ding{109} & \ding{108} & G \\
\cite{huang2021characteristics} & 2021 & \ding{109} & \ding{109} & \ding{109} & \ding{109} & O \\
\cite{wood2021gendered} & 2021 & \ding{108} & \ding{109} & \ding{108} & \ding{108} & G \\
\cite{Lu2022} & 2022 & \ding{109} & \ding{109} & \ding{109} & \ding{109} & S \\
\bottomrule
\end{tabular}
\end{minipage}
\end{table*}

In this section we discuss how human mobility research papers dealt with some of the ethical aspects presented in the previous section. We present numbers based on a review of \Npapers{} papers from CS and general sciences, published from 1998 and 2022, which used individual location data, totalling \Ndatasets{} datasets (\textit{i.e.}, papers could contain multiple sets). In all cases, we consider only the first publication to include a given dataset (see \S~\ref{sec:scales}), and a basic of description of all papers is presented in Table~\ref{tab:summary}.

We begin by making a distinction between datasets collected \emph{deliberately} for a study and those which were shared with researchers through an \emph{agreement}. From all datasets, one in four (\Nstudydata{}, or \Pstudydata{}\%) were collected \emph{deliberately}, as an integral part of the study. Additionally, sets covering larger areas, or that were collected more recently, or sets with larger subject counts showed a lower likelihood of a deliberate collection effort. This could, in part, be explained by a predominant use of CDR and OSN/Web based data in recent years, as depicted in Figure~\ref{fig:scaletime}, as those sources often capture location data for basic operations, such as billing, and which are then repurposed for mobility research.

Data sharing remains the exception, understandably. While data sharing has grown in importance for reproducing and validating existing results or fostering new research, privacy concerns and NDA's often prevent any data being shared. From all sets, one in ten (\Nshared{}, or \Pshared\%) were shared in either their original form or modified along with the paper (\textit{e.g.}, \cite{kotz2005crawdad,Gianni}). However, we note that, in some cases, authors provide the contact of the data provider stating that those sets could still be retrieved upon agreement.

In some cases, papers took deliberate actions to protect subjects privacy, beyond anonymizing user identifiers one single time. For that, we identified one in 15 (\Npreproc{}, or \Ppreproc{}\%) of sets received extra steps, with no clear distinction between CS and general sciences. Additionally, we note that such measures became more common in more recent papers. Examples of such mitigating measures include \emph{spatial aggregation} (\textit{e.g.}, \cite{Becker2011,lu2013approaching,alessandretti2018evidence,kraemer2020mapping}) where records per individual are still available but the spatial resolution is reduced, similarly with \emph{temporal aggregation} (\textit{e.g.}, \cite{Lutu2020}) where records are grouped in large time slices, and \emph{periodical changes in user identifiers} (\textit{e.g.}, \cite{Heiler2020}). It is worth noting that only some papers state what might be clear, that
``\textit{obeying data privacy regulations is important when analyzing mobile phone usage data}'',
and after explaining what was done conclude that
``\textit{thereby the local regulations have been met and the recommendations of the GSMA, the alliance of mobile phone providers have been followed}.''~\cite{Heiler2020}

As discussed previously, beyond the use of user identifiers, subjects can be de-anonymized through other bits of information, such as the places they visit, the schedule they follow, and their social network. Therefore, approaches can be taken to minimize the availability of such information, which in many cases still allows the same research questions to be answer. One of these approaches is a periodic change of anonymized identifiers per subject (\textit{e.g.}, daily), which was observed in only one in 16 (\NuidChanged, or \PuidChanged{}\%) of the datasets we observed. Knowing the home of a subject (or approximate location) can provide unicity to a subject and in six out of seven (\Nhome{}, or \Phome{}\%) of sets this information was present but only in a third (\Nhomeused{}, or \Phomeused{}\%) of those that information was actually used. Similarly, information on other places, schedule and social network was available in \Nplaces{} (\Pplaces{}\%), \Nschedule{} (\Pschedule{}\%) and \Nsocial{} (\Psocial{}\%) of datasets respectively, out of which were only used in \Nplacesused{} (\Pplacesused{}\%), \Nscheduleused{} (\Pscheduleused{}\%) and \Nsocialused{} (\Psocialused{}\%) respectively. These numbers highlight the availability of information past mobility research had access to which was not needed in some way, and therefore could have been left out during the data collection process. These extra cautionary steps taken in future research may help with public opinion and trust.

Given all the extra data available to researchers, any form of pre-processing beyond the change in identifiers is our next observation. For that, only one in 15 (\Npreproc{}, or \Ppreproc{}\%) of sets went through some kind of pre-processing done, out of which three out of five (\Npreprocmeth{}, or \Ppreprocmeth{}\%) had their methods clearly explained and nine in ten (\Npreprocwho{}, or \Ppreprocwho{}\%) clearly stated \emph{who} performed those extra steps. Once again, these extra bits of information help not only other researchers better understand what was done, but more importantly help assure public trust on the true aims and practices of mobility research.

Another important aspect of this analysis is how papers dealt with ethics. Even though ERB/IRB are not mandatory in most of the outlets we studied, one in six (\NIRB{}, or \PIRB{}\%) of papers had an IRB statement. Out of these studies, only one (\cite{luo2017inferring}) provided a protocol which allows for an independent verification of that approval.

\paragraph{Informed consent}
Given the importance of informed consent previously discussed (\S~\ref{sec:ethics}), we report that one in eight (\NethConsent, or \PethConsent{}\%) of the listed papers declared having received authorization from subjects to perform the study being presented. This goes with how people expect their data being used in accordance with the terms and services they sign before using a service, that is, an OSN user will agree in having their data being stored and processed to be used by the platform providing them the services, not some third party trying to understand changes in mobility.
In most cases when consent was given, papers describe consent being given for a particular study, \textit{e.g.}, ``\textit{All users of Locaccino, regardless of how they were recruited, gave informed consent to participate in the study work}''~\cite{Cranshaw2010}. There are, however, exceptional cases in which a double-consent was requested to cover not only the data collection but also the specifics of the study, \textit{e.g.}, ``\textit{Therefore, YJ performed a double consent process, where the users who have given their consent to the usage of location information and web search queries were asked again, if they wish to provide their consent to be included in the [COVID19] dataset.}''~\cite{Yabe2022}.

\section{Outlook on Human Mobility Research Using Individual Data} \label{sec:outlook}

In this paper, we surveyed \Npapers{} papers dating back over 20 years with the goal of reviewing how individual mobility data were used, and how methods and results were communicated with special attention to privacy and ethics. We now conclude with a series of final remarks, pushing for a future with more accountability, protecting the trust of subjects, researchers and the general public to avoid a \emph{tragedy of the data commons}~\cite{yakowitz2011tragedy}. Given all the data and methods reviewed in this paper, we summarize a series of ideas and guidelines networking research using personal data could follow in the upcoming years. We do not focus on the exact steps to achieve those goals, but we leave those to the community to decide.

\smallskip
\noindent\textbf{IMC ethical considerations.}
Ethical concerns have been part of network measurement community (especially IMC) for long~\cite{Dietrich2014Ethics}.
Ethical statements are required by several scientific outlets, such as those under SIGCOMM~\cite{gotterbarn2018acm}, typically in the form of a required section where ethical considerations are discussed~\cite{Partridge2016}, and more recently, requiring that the IRB review application form be submitted along with the manuscript.
Furthermore, the inclusion of ethics course as an integral part of Computer Science curricula have also been extensively discussed over the years~\cite{force2020computing,saltz2018key}, fostering a growing ethical ethos among scientists.
Therefore, we believe the mindset of ethically doing research with personal data should be present from the initial steps of any investigation endeavour, similar to the moral philosophy discussed by Bietti~\cite{bietti2020ethics} for the tech industry. However, as we have highlighted with our review, a significant number of papers do not include ethical discussions even amongst recent work, let alone consider IRB reviews which need to happen \textit{a priori} (\textit{i.e.}, need to be considered even before a submission, not \emph{only once this is accepted}). 
We also find in our review that the definitions of \enquote{ethics} and \enquote{personal data} have evolved over time. Methods to ask for consent have become more minimalist (and often implicit) so as to not inhibit a user's \enquote{quality-of-experience}.
Simultaneously, we observe that the ethical boundaries of large-scale (or in-the-wild) measurements are only known implicitly to researchers already well-versed with the field.  
Lets take, for example, research in conducting Internet-wide scans.
While there is plethora of work easily accessible via quick search on tools and methods to perform IP scanning~\cite{scan-tools}; reports on best scanning practices~\cite{dittrich2012menlo}, ownership within the context of Internet~\cite{Dietrich2014Ethics} and guidelines for designing ethical scanning methodologies~\cite{durumeric2013zmap} are much less popularly known and sought after. 
Additionally, we believe a minimalist and careful handling of personal data must be practiced at all steps of the scientific process. This cautionary approach, combined with higher clarity in describing how data were used, and higher accountability are of paramount relevance for a clearer implementation of ethics in research.
That is, a call for the wider adoption of the intrinsic values of ethics over the instrumental ones~\cite{bietti2020ethics}, where in the former ethics is seen as a commitment to a process while in the latter ethics is simply a means to an end (\textit{e.g.}, fulfill requirements in a document being written).
IRB reviews should be followed, and if they fail, then a redesign of the study is needed. Conferences should require a copy of the IRB check, as it is already done at IMC and other outlets. 

\smallskip
\noindent\textbf{Scientific Data Sharing}
Amongst the issues discussed, we believe this is one the hardest to meaningfully address with assertive measures. While on the one hand data sharing is of significant importance to the validation and reproducibility of research~\cite{Demir2022}, on the other hand protecting the privacy of subjects prevents sets being shared in their original form. Current solutions require at least some level of differential privacy (see~\ref{subsec:anon}) to be met, which results in loss of information, or sharing some kind of synthetic dataset, which may still hold valuable information about the original subjects or render significantly different results from those present in the original work. Restricting research to use only shareable data are likely not the solution either. That is, limiting papers from being published only with data that have been fully anonymized and which could be shared, even under any strong NDA, cannot be a solution as it may hinder scientific development. To the problem of sharing data while protecting subject's privacy, no single rule shall be applied and the work of ethical review boards, once more, becomes even more relevant.

\smallskip
\noindent\textbf{Ethical statements}
A comprehensive and thorough ethical statement must be present in all research done using individual data. While in some venues this is the norm, there are still outlets which do not enforce this as a requirement. Note, however, that is not an invitation for the inception of ``ethics wash''~\cite{bietti2020ethics} (\textit{i.e.,} whitewash) in our community, where statements will be written only to fulfill a given requirement using \textit{boilerplate} sentences, hiding the fact that no particular attention was paid to ethics or privacy. That is, ethical statements should not be used to cover up for bad behavior or any form of wrongdoing.

\smallskip
\noindent\textbf{The explicit consent dilemma}
As we have shown, publication are using larger pools of subjects and using a growing variety of data sources. Therefore, it is na\"{i}ve to expect studies with hundreds of millions of subjects will get explicit consent from every single subject, when for example, not all questions are known \textit{a priori}, as previously discussed in the Menlo Report~\cite{Bailey2012Menlo}. While no alternatives for better data governance are put in place it is the researcher's responsibility to take all cautionary steps to ensure the privacy of subjects is preserved, that all the steps taken to process and analyze those data are clearly described, and that the benefits of their research are clearly communicated for all audiences.

\smallskip
\noindent\textbf{Ethics check committee at conferences and journals}
Similar to the already existing reproducibility checks, conferences and journals could implement an ethics check committee, responsible for double-checking whether essential ethics guidelines were followed (\textit{e.g.}, ethical statement, IRB approval, informed consent). This would be available for any paper containing personal data, and as its reproducibility counterpart, would not affect the acceptance likelihood of the work. We hypothesize this \emph{opt-in} solution could create incentives for authors to behave more ethically, and make that clear on their manuscripts, without stringent measures that could hinder scientific development in the short term.

\smallskip
\noindent\textbf{A model for better individual mobility data governance}
In order to get around some of the issues discussed in this paper, we believe better traceability and accountability are plausible alternatives, but we leave the exact details of implementation for the community to decide. Regardless of the architecture of such solution, this system would be responsible for collecting, filtering, aggregating and sharing these data with internet services interested in using that information. These data points could include readings from sensors (accelerometers, GPS), contextual physical information such as proximity to others through Bluetooth, as well as online behavior. Users would then have the choice of whether or not any of those data are collected, how and if they should be aggregated or filtered, before being shared. Researchers could, in turn, publish calls for data, explaining in details their objectives and data are needed. This system could also work as a bookkeeper to all accesses to a subject's data, as well as whenever the original data are finally deleted, ensuring better accountability and possibly wider availability of data for research studies.